\documentclass[12pt,preprint]{aastex}

\newcommand{\Hnot}{70\,km\,s$^{-1}$\,Mpc$^{-1}$}
\newcommand{\Ol}{\Omega_\Lambda}
\newcommand{\Om}{\Omega_m}
\newcommand{\ndetect}{71}         
\newcommand{\ndust}{53}\newcommand{\percentdust}{60}
\newcommand{\ntotal}{88}          
\newcommand{\nletter}{27}         
\newcommand{\Lsun}{L$_\odot$}
\newcommand{\zmax}{1.2}
\newcommand{\zmaxsource}{3C324}

\slugcomment{accepted for publication in ApJ Letter }

\shorttitle{Hot Dust in 3C Galaxies}
\shortauthors{}

\begin{document}

\title{Hot Dust in Radio-Loud Active Galactic Nuclei}

\author{Wolfram Freudling}
\affil{Space Telescope -- European Coordinating Facility \\
       European Southern Observatory \\
       Karl-Schwarzschild-Str. 2 \\
       85748 Garching \\
       Germany }
\email{wfreudli@eso.org}

\and

\author{Ralf Siebenmorgen}
\affil{ European Southern Observatory \\
       Karl-Schwarzschild-Str. 2 \\
       85748 Garching \\
       Germany }
\email{rsiebenm@eso.org}

\begin{abstract}   
We have measured mid-infrared (MIR) fluxes of 3C sources on images
taken with ISOCAM onboard the {\it Infrared Space Observatory (ISO)}.
The photometric data were combined with existing photometry at other
wavelengths to assemble the spectral energy distribution (SED) for
each galaxy from infrared to radio wavelengths. In addition, we used
ISOPHOT spectra to compute average MIR spectra for different types of
Active Galactic Nuclei (AGNs).  The MIR emission of \ndust\ of our
sources shows evidence of dust.  We find a clear correlation between
the SEDs, emission from polycyclic aromatic hydrocarbons (PAHs), and
the type of AGN. Specifically, we find that in broad emission line
radio galaxies (BLRGs), the dust emission peaks at
$\lambda\approx40\mu$m and PAH bands are weak, whereas emission in
comparable narrow line radio galaxies (NLRGs) peaks at a longer
wavelength of $\lambda\approx100\mu$m and the PAH bands are much
stronger. Although less pronounced, a similar trend is seen when
comparing quasars (QSOs) with high-luminosity NLRGs.  We used dust
radiative transfer models with a central heating source to describe
the SEDs. The difference in the dust emission appears to be the effect
of heating by radiation from the central engine.  In type~1 AGNs
(BLRGs and QSOs), the broad line region (BLR) is directly visible and
therefore hot ($T\gtrsim300K$) dust dominates the emission. In this
region, PAHs are destroyed which explains the weak PAH emission. On
the other hand, the BLR is hidden in type~2 AGNs (NLRGs). Their SEDs
are therefore dominated by cooler dust including PAHs.
\end{abstract}


\section{Introduction}

The widely accepted paradigm for the power source of AGNs is a massive
black hole. In this picture, accretion of dust and gas powers the
AGN. Beamed radiation escapes from the central engine and affects the
surrounding gas and dust.  The effect on gas can directly be observed
in close-by AGNs through the spatial separation of ionized and neutral
gas (e.g. Tadhunter and Tsvetanov 1989, Prieto and Freudling
1993). The impact of this radiation on dust, if present, is expected
to be similarly dramatic. The intense radiation field in the immediate
vicinity of the central engine prevents dust from forming and
evaporates any dust approaching that region. In this region, PAHs
cannot exist. At somewhat larger distances from the nucleus, the
radiation field heats the dust to just below its evaporation
temperature. Dust at such distances will have temperatures of about
1500K. Cold dust ($T\lesssim50K$) should exist only at distances
greater than several hundred pc to a few kpc. At intermediate
distances, a region should exist were large quantities of dust are
significantly heated by the AGN to temperatures of a few hundred
Kelvin and should therefore be detectable at MIR wavelengths.

Unfortunately, detailed mapping of dust temperatures in AGN host
galaxies is not feasible at present. Our knowledge of dust in 3C
sources is therefore based on low spatial resolution photometry.
Observations with the {\it Infrared Astronomical Satellite (IRAS)}
showed that 3C sources which are bright at $\lambda=60\mu$m are
typically much fainter or not detectable at
$\lambda=12\mu$m \citep[][and references therein]{heck}. This
signature of cold dust is also seen in the more recent data presented
by \cite{Meise}. \citet{heck} find that the NLRGs are about a factor
of 4 to 5 fainter in the MIR than BLRGs, but are similar to BLRGs at
$\lambda=60\mu$m. These differences are qualitatively consistent with
unification schemes since hotter dust is visible in BLRGs due to their
viewing angle.

ISOCAM was more than an order of magnitude more sensitive than {\sl IRAS}
at MIR wavelengths. It was able to detect the hot dust
component in 3C sources at much higher redshifts and therefore expand
comparison of type~1 and~2 AGNs to a much wider range of luminosities.
In this letter, we report first results from a program to investigate
the hot dust component in 3C sources based on ISOCAM observations.
A detailed discussion of our data reduction and modelling procedures
as well as a full presentation of our data are deferred to a later
paper
\citep[][hereafter SFKH]{mainpaper}.

\section{Observations and Data Reduction}

\subsection{The Sample}

The goal of our program is to investigate dust emission in AGNs, and
in particular to search for differences related to the orientation of
the central engine.  Constraining the SED emission of both hot and
cold dust requires infrared photometry covering the full wavelength
range from $\mu$m to mm wavelengths.  We used the 3CR catalogue as our
parent catalogue because its 178 MHz selection should be essentially
unbiased in its inclusion of AGNs at different viewing angles. In
addition, photometric measurements at far infrared (FIR) and mm
wavelengths are available for large subsamples of the catalogue
\citep[e.g.][]{Meise,Haas}. By complementing these available data with
our new MIR measurements, we were able to construct SEDs from optical to
mm wavelengths.

We have compiled our sample by cross-correlating the
position of 3C sources with the list of ISOCAM observations available
from the ISO post mission archive. We
then extracted all ISOCAM images which potentially include 3C
sources. Each reduced image was visually inspected and the pointing of
the spacecraft was verified using stellar and other sources contained
in the images.  We report here on that subsample of
our data for which the combined ISOCAM photometry, data from the
literature and ISOPHOT photometry by \citet{Haas} suffice to constrain
the full infrared SED of our sources.

\subsection{Data Reduction}

ISOCAM images in the ISO archive have been obtained for a variety of
programs and purposes and therefore vary in depth, resolution, used
wavelength band and dither strategy. We re-processed the raw data
using a common set of procedures aimed at obtaining the best possible
calibration and homogeneity of our photometric measurements.  We
followed the ISOCAM handbook \citep{isohandbook} and used the CIA
software \citep{cia} to subtract darks, remove cosmic rays hits,
remove the effect of flux transients, and finally flatfield, re-sample
and co-add the individual exposures. The typical useful field of view
of the images is about $(1.5 \rm{arcmin})^2$ sampled with $(3
\rm{arcsec})^2$ pixels. 
Most of the detected 3C sources were not or only barely spatially
resolved by ISOCAM. In those cases, apertures with a radius of 10
arcsec were used, and aperture losses were modelled using synthetic
point spread functions \citep{PSF}. For extended sources, we chose
apertures which measure the spatially integrated flux of the source.

For most 3C sources, photometric information at optical, near-infrared
and/or far-infrared, and mm wavelengths is available. We compiled SEDs
between 0.5 and 1300 $\mu$m for all sources in our sample using data
listed in the {\it NASA Extragalactic Database (NED)}, the ISOPHOT
measurement given in \citet{Haas} and other recent papers (see
SFKH). In compiling the SEDs, we only used photometric data which
includes, like our own ISOCAM photometry, the integrated flux from the
whole host galaxy. For each SED, we estimated the contribution of
synchrotron radiation to the MIR fluxes by extrapolating the radio core
flux.  We found that the synchrotron radiation is a negligible
contribution to the MIR flux for all SEDs used in this paper.

\section{Results}

We retrieved ISOCAM images for a total of \ntotal\ sources. The rms
noise in the reduced images ranges from 0.5 to 5 mJy.
We detected a total of \ndetect\ galaxies.  The high detection rate 
indicates that hot dust is common in radio-loud AGNs, and that 3C
sources are bright enough in the MIR so that they will be readily
accessible to detailed studies by future instruments such as SIRTF.

A steep rise of the flux from optical wavelengths below 1$\mu$m to the
MIR indicates the presence of a significant amount of dust at
$T\approx300$K for a large fraction of our sample.  In order to
estimate the contribution of stars to the MIR, we fitted 4000K black
body spectra to the short wavelength part of the SEDs. This
temperature  deliberately was taken to be on the low side of the
range of possible average stellar temperatures so that the MIR
emission attributed to stars is an upper limit. Taking this stellar
contribution into account, we found clear evidence of dust in \ndust\
sources, i.e. \percentdust\% of our initial sample. The highest
redshift source with a clear detection of hot dust is \zmaxsource\ at
a redshift of \zmax.  For a total of \nletter\ sources, our data
suffice to construct SEDs.  The measured ISO fluxes for them are
presented in table~1.

\section{SED Modelling}

Interpretation of the IR spectra requires detailed knowledge of the
energy sources and the distribution and composition of gas and dust
which absorbs and re-transmits the radiation. In the case of the AGNs
discussed in this paper, the primary energy source is the central
engine in the nucleus enshrouded by dust clouds. Dust close to the
nucleus absorbs the radiation from the nucleus and re-radiates
it. This emission depends mostly on the total power and spectrum of
the central engine, and the distribution and properties of dust. While
the composition of the dust can reasonably assumed to be similar to
local dust and is therefore well constrained, its distribution around
the power source of the AGN is largely unknown.  It is generally
assumed that the geometrical distribution of dust around AGNs is
similar to a torus or warped disk \citep[e.g.,][]{review}.  However,
observations of dust features close to the nucleus of AGNs with the
HST reveal a variety of morphologies \citep{Mat2}.  These observations
and the fact that dust at distance much larger than the size of the
putative torus shapes the MIR spectrum \citep{Far} motivated us to
adopt a simple approach to the modelling of dust SEDs in AGNs. Instead
of choosing between possible geometrical configurations, we carried
out spherical radiative transfer calculations with a central heating
source.  In this approach, the size of the model sphere depends on
the viewing angle because of the intrinsic asymmetry of AGNs. We used
a power law $L \propto\nu^{-0.7}$ between $\lambda=10$\AA\ and $2\mu$m
as the input spectrum of the central sources. The exciting radiation
field is therefore significantly harder than for star bursts leading
to stronger photo-destruction of small grains \citep{Timmi}.  Had we
used a more distributed source of energy within the nucleus, the
predicted PAH emission would have been much stronger. \label{model}

We numerically solved the radiative transfer equations for a grid of
models, varying the total luminosity of the central engine, the total
extinction and the outer radius of the dust clouds. The dust in our
model consists of carbon grains, silicate grains and PAHs. The initial
density within the sphere is constant, but dust evaporation and PAH
destruction are taken into account in a self-consistent way. We
computed model SEDs for outer radii ranging from 0.125 to 16kpc,
luminosities ranging from $10^{9}$ to $10^{14}$\Lsun, and visual
extinction ranging from 1 to 128 magnitudes. The full set of models is
available at {\tt http://eso.org/\~\ rsiebenm/agn\_models.html}.  The
parameters of the best fitting model for each observed SED are listed
in table~1. In our modelling, all emission from the central engine is
reprocessed by dust. The input power to our models is therefore
identical to the total dust luminosity, which we use below to divide
our sample into subsamples.

\section{Discussion}

\subsection{Averaged SEDs}

In order to investigate differences in dust emission for AGNs of
type~1 and~2, we divided our sample into four different subsamples.
Type~1 AGNs were subdivided into BLRGs and QSOs. Type~2 AGNs with dust
luminosities fainter then $10^{11}$\Lsun were assigned to a low
luminosity NRLG subsample, and those which are as bright or brighter
than this limit to an equivalent high luminosity subsample. This
luminosity limit was chosen so that the average dust luminosities of the
high, respectively low luminosity NLRGs resemble those of the QSOs,
respective BLRGs. This subdivision also led to samples comparable in
redshift and extended radio emission. Specifically, the mean redshift
and extended radio luminosity of the high-luminosity NLRGs
($z=0.53\pm0.4$ and $L_r=10^{27.5\pm 0.6}$W/Hz/sr) match those of the
QSOs ($z=0.53\pm0.2$ and $L_r=10^{27.1\pm 1.1}$W/Hz/sr), and the
redshifts and radio luminosities of the low luminosity NLRGs
($z=0.06\pm0.06$ and $L_r=10^{25.3\pm 0.7}$W/Hz/sr) resemble those of
the BLRGs ($z=0.11\pm0.06$ and $L_r=10^{25.9\pm 0.4}$W/Hz/sr).

\begin{figure}
\plotone{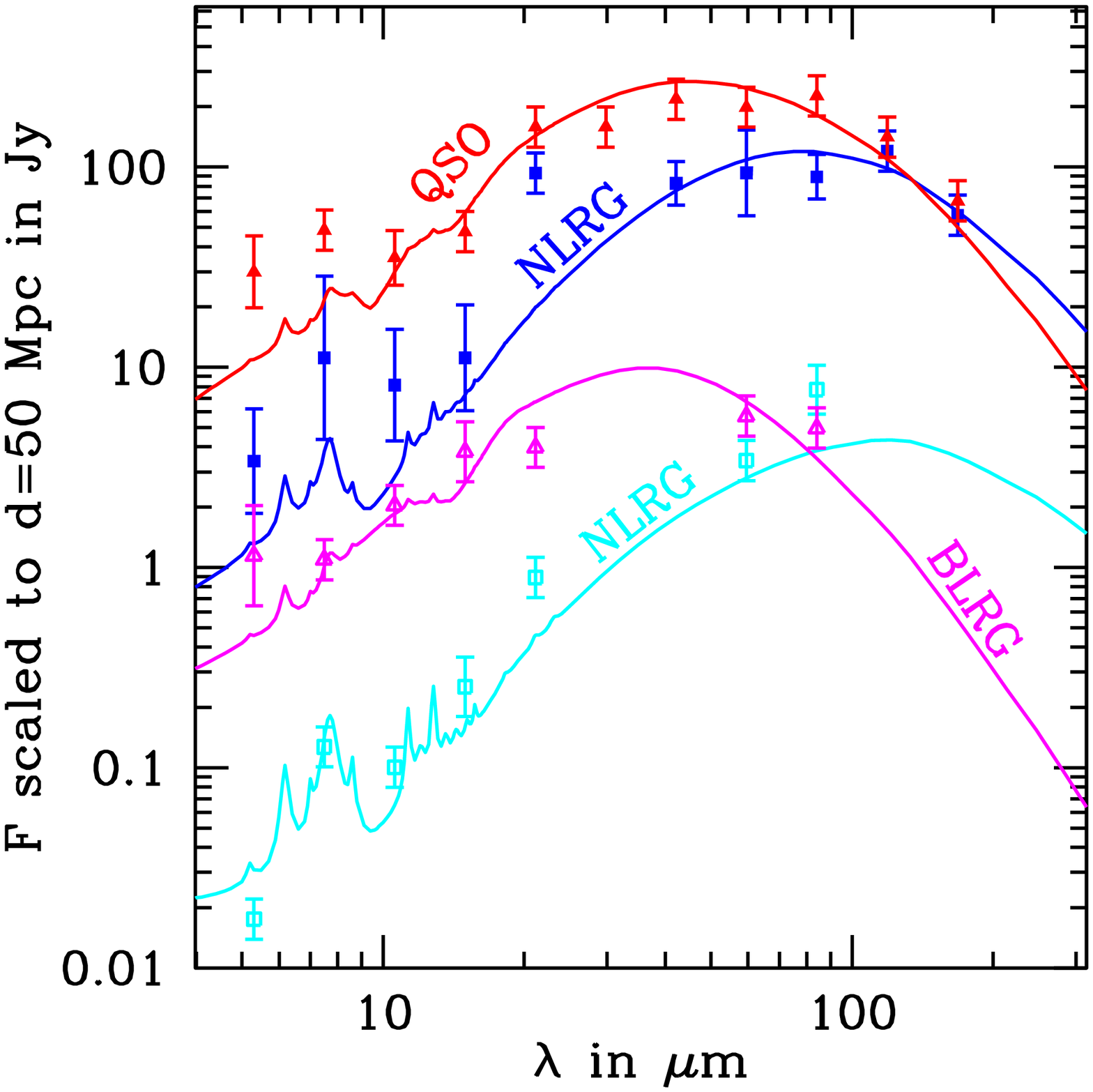}
\caption{
Mean SEDs for different AGN types.  Filled triangle represent QSOs,
filled squares NLRGs with $L_{\rm dust} > 10^{11}$\Lsun (black lines),
open triangles show BLRGs and open squares show NLRGs with $L_{\rm
dust} < 10^{11}$\Lsun. Error bars are the 1$\sigma$ uncertainty of the
mean in each wavelength bin. The solid lines are the corresponding
averaged model for each AGN type. }\label{SED}
\end{figure}

We computed logarithmically averaged SEDs for each subsample to search for
differences in their shapes. Specifically, the procedure we used was
the following.  First, we redshifted each observed SED to the
restframe, and scaled it to the same interpolated flux at
$\lambda=25\mu$m. Next, we computed logarithmic flux averages in
wavelength bins.  The averaged SEDs were then scaled so that their
integrated fluxes equal the average of the integrated fluxes of all
corresponding spectra.  The same procedure without the wavelength
binning was used on the model SEDs for each contributing source.  The
resulting averaged observed and model SEDs are shown in
Figure~\ref{SED}.  We have also used  different averaging procedures
and found that they did not alter the conclusions presented in this
paper.

Figure~\ref{SED} shows some clear and highly significant differences in
the average SEDs of different AGN types. The dust emission peaks at
$\lambda\approx40$ to $50\mu$m for QSOs and BLRGs, and around 70 to
100$\mu$m for NLRGs. This difference in the peak wavelength by a
factor of about 2 between type~1 and~2 objects seems to be virtually
independent of luminosity. The difference reflects different
temperature distributions of large dust grains and  can be
understood within the framework of unified models. For type~1 AGNs
where the BLR is visible, unobscured hot dust dominates the SED. By
contrast, the NLRG's large extinction
obscures the BLR and pre-dominantly colder dust is visible.

\subsection{Strength of PAH bands}

A more specific test of the heating mechanism is the strength of PAH
bands. As discussed in section~\ref{model}, the PAH strength is a
sensitive indicator of the heating source because PAHs are destroyed by
hard ionizing radiation. Despite the low spectral resolution of the
SEDs shown in figure~\ref{SED}, some indications of the PAH bands are
visible in the SEDs of the type~2 AGNs, but not for the type~1 AGNs. 
This observation seems to confirm the destruction of PAHs near
the central heating source.

\begin{figure}
\plotone{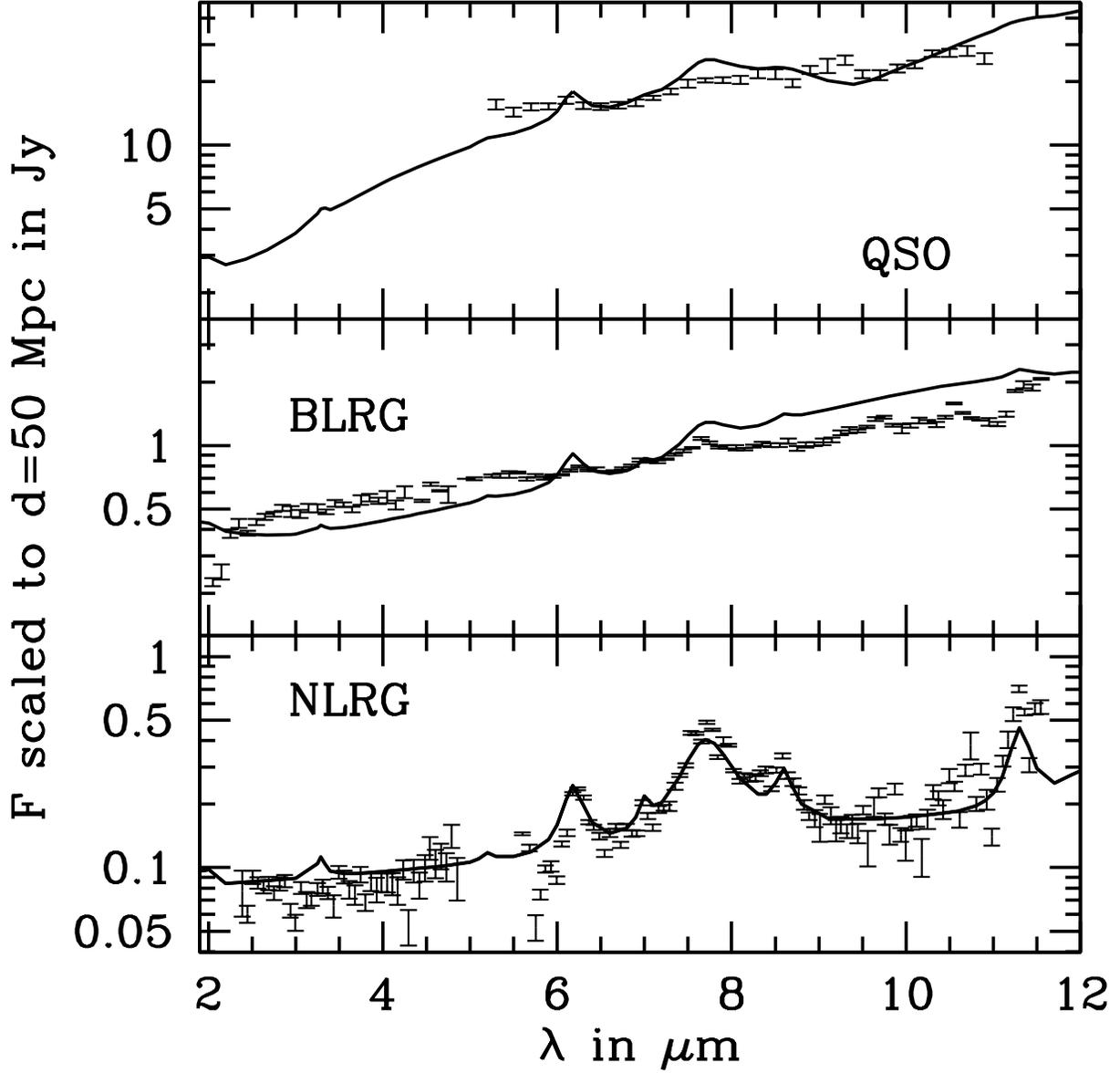}
\caption{{\it Upper panel:} mean model spectrum of QSOs (solid line)
in our sample superimposed on average ISOPHOT spectrum of PG quasars
(error bars). {\it Middle panel:} mean model of BLRGs 
superimposed on observed average ISOPHOT spectra (error bars).  {\it
Lower panel:} mean model SED of NLRGs with $L_{\rm dust} <
10^{11}$\Lsun (solid line) superimposed on observed average Seyfert 2
spectra (error bars).}\label{MIR}
\end{figure}

We used ISOPHOT spectra of AGNs to investigate the strength of PAH
bands as a function of AGN type in more detail. For that purpose, we
assembled average spectra for each AGN type. Since radio loud and
radio quiet AGNs have similar IR spectra \citep[e.g.][]{irqso}, we
included both kinds of AGNs in those averages.  We used the spectra
presented by \citet{Haas_PG} to compute an average QSO spectrum.  For
the BLRGs, we used ISOPHOT spectra of 3C sources available from the
ISO archive and the Seyfert~I spectra from \citet{spectra} .  Finally,
we used the average spectrum of Seyfert~II galaxies presented by
\citet{spectra} as our average type~1 AGN spectrum.  The flux of each
averaged spectrum was scaled to the average of the corresponding AGN
type in our 3C sample.  These scaled spectra are presented in
Figure~\ref{MIR}.

The spectral index $\alpha$ was estimated by fitting $F_\lambda
\propto \lambda^{-\alpha}$ to the continuum of the averaged spectra.
We found $\alpha=-0.91\pm0.06$ for the QSOs, $\alpha=-0.70\pm0.03$ for
the BLRGs and $\alpha=-0.76\pm0.05$ for the Seyfert~IIs.  We note that
due to the similarity between the 3C spectra of type 1 and Seyfert~I
spectra, the continuum slope as well as the PAH strength in our
average spectrum of the BLRGs is indistinguishable from that of the
average spectrum of Seyfert~I galaxies shown in Figure~7 of
\citet{spectra}.

A comparison of the three spectra in Figure~\ref{MIR} shows a clear
difference between the PAH bands in spectra of AGNs of similar
luminosity but different type.  PAH bands in the spectra of type 1
AGNs are more than a factor of three stronger than in the
corresponding type 2 spectra.

For comparison, we plotted the SED models shown in Figure~\ref{SED}
for the QSO, BLRG and low-luminosity NRLG on top of the spectra in
Figure~\ref{MIR}. One can immediately recognize a basic correspondence
between the continuum shapes and PAH strength of the models with the
spectra. The fact that models fitted to the overall SEDs correctly
predict the strength of the PAH bands in independent samples of the
same AGNs type suggests a basic similarity in the physics within each
of the different types of AGNs. 

The relatively weak PAH bands in type~1 AGNs can be understood in
terms of the unified models. In type~1 spectra, a large fraction of
the visible flux originates from the BLR region where only few PAHs
survive. This region is obscured in type~2 AGNs and their spectra are
hence dominated by cooler dust located at larger distances from the
nucleus where PAHs are shielded from the photo-destruction. Type~1
and~2 AGNs spectra differ because of the different relative
contribution of hot dust to the total emission. When comparing
spectra of AGNs with the same total {\em dust} luminosity, the different
relative contribution of cold dust leads 
to the observed differences in the total PAH emission.

\section{Summary and Conclusion}

We have detected hot dust in ISOCAM images of 3C galaxies out to
redshift z$\approx$\zmax.  We found a clear correlation between the
type of the AGN and the overall SEDs in the sense that SEDs of BLRG
and QSOs are dominated by emission from dust which is about a factor
of 2 hotter than the one in comparable NRLGs.  MIR spectra show that
PAH emission bands in spectra of type~1 AGNs are significantly weaker
than the ones in spectra of type~2 AGNs with the same total dust
luminosity. These trends are naturally predicted if dust in AGNs is
predominantly heated by hard radiation from a central engine which
destroys PAHs close to the nucleus and heats larger dust grains at
intermediate distances.

\acknowledgments

{\it Acknowledgments}. We thank Martin Haas and Bernhard Schulz for providing digital versions of their ISOPHOT spectra.

\clearpage

\clearpage

\begin{deluxetable}{ccccccc}
\tablecolumns{7}
\tablewidth{0pc}
\tablecaption{ISOCAM observations of 3C sources}
\tablehead{
\colhead{Name} &\colhead{Type} & \colhead{ $F_{\rm ISOCAM}$ } & Filter\tablenotemark{a} &\colhead{ $R$} &\colhead{ $log(L)$\tablenotemark{b}}& \colhead{$A_V$} \\
 & \colhead{} & \colhead{mJy} &  & \colhead{kpc} & \colhead{\Lsun} & \colhead{mag.}}
\startdata
 3C006.1         &   NLRG        & 0.9  $\pm$ 0.2 & 10 & 16    &  11.75 &  32  \\
 3C020           &   NLRG        & 5.3  $\pm$ 0.4 & 10 & 2     &  11.25 &  16  \\
 3C031           &   LERG        & 25   $\pm$ 1   & 10 & 0.125 &  10.00 &  2   \\
 3C033.1         &   BLRG        & 18   $\pm$ 1   & 10 & 0.125 &  11.25 &  16  \\
 3C048           &   QSO         & 59   $\pm$ 4   & 3  & 4     &  12.75 &  32  \\
 3C061.1         &   NLRG        & 0.9  $\pm$ 0.2 & 10 & 1     &  10.75 &  32  \\
 3C071           &   NLRG        & 50830$\pm$ 2540& 3  & 2     &  11.50 &  16  \\
 3C079           &   NLRG        & 21   $\pm$ 2   & 10 & 0.5   &  11.75 &  16  \\
 3C084           &   NLRG        & 230  $\pm$ 20  & 2  & 2     &  11.25 &  4   \\
 3C098           &   NLRG        & 24   $\pm$ 2   & 3  & 0.125 &  10.00 &  32  \\
 3C249.1         &   QSO         & 19   $\pm$ 1   & 10 & 0.125 &  11.75 &  32  \\
 3C277.3         &   BLRG        & 8.5  $\pm$ 0.5 & 3  & 0.25  &  10.50 &  1   \\
 3C293           &   LERG        & 19   $\pm$ 2   & 10 & 8     &  10.75 &  2   \\
 3C305           &   NLRG        & 21   $\pm$ 1   & 10 & 8     &  10.75 &  1   \\
 3C309.1         &   QSO         & 8.2  $\pm$ 0.5 & 10 & 0.25  &  12.75 &  32  \\
 3C321           &   NLRG        & 27   $\pm$  1  & 10 & 2     &  11.50 &  16  \\
 3C324           &   NLRG        & 1.7  $\pm$ 0.2 & 10 & 4     &  12.25 &  8   \\
 3C330           &   NLRG        & 1.7  $\pm$ 0.2 & 10 & 0.125 &  11.75 &  32  \\
 3C338           &   NLRG        & 8.8  $\pm$ 0.5 & 10 & 16    &  10.25 &  16  \\
 3C351           &   QSO         & 32   $\pm$ 2   & 10 & 0.5   &  12.25 &  32  \\
 3C356           &   NLRG        & 0.6  $\pm$ 0.2 & 10 & 4     &  12.25 &  64  \\
 3C380           &   QSO         & 14   $\pm$ 1   & 10 & 1     &  12.50 &  32  \\
 3C381           &   BLRG        & 19   $\pm$ 1   & 10 & 0.25  &  11.25 &  32  \\
 3C382           &   BLRG        & 85   $\pm$ 5   & 10 & 0.125 &  11.00 &  2   \\
 3C390.3         &   BLRG        & 94   $\pm$ 5   & 10 & 0.125 &  11.00 &  4   \\
 3C445           &   BLRG        & 210  $\pm$ 11  & 3  & 0.125 &  11.50 &  16  \\
 3C459           &   NLRG        & 29.5 $\pm$  4  & 3  & 2     &  12.25 &  128 \\
\enddata
\tablenotetext{a}{ISOCAM Filters encoded as  2=LW2 ($\lambda=6.7\mu$m),
                                            10=LW10 ($\lambda=12.0\mu$m),
                                             3=LW3 ($\lambda=14.3\mu$m) }
\tablenotetext{b}{based on $H_o$=\Hnot, $\Ol$=0.70, and $\Om=0.3$}
\end{deluxetable}

\end{document}